\documentclass{amsart}

\usepackage{amssymb,amsfonts,latexsym,amsmath,amsthm,wrapfig,graphicx, hyperref, 
}

\usepackage{bm}
\usepackage{epsfig}

\input xy
\xyoption{all}

\numberwithin{equation}{section}
\newtheorem{thm}{Theorem}[section]

\newtheorem{lemma}[thm]{Lemma}

\newtheorem{definition}[thm]{Definition}
\newtheorem{conjecture}{Conjecture}
\newtheorem*{question}{Question}

\newcommand{\RR}{\mathbb{R}}
\newcommand{\R}{\mathbb{R}}
\newcommand{\CC}{\mathbb{C}}

\newcommand{\ZZ}{\mathbb{Z}}

\newcommand{\MM}{\mathbb{M}}

\newcommand{\supp}{\rm{supp}}
\newcommand{\cp}{\rm{cap}}

\newcommand{\emk}{{\rm{cap}}}

\newcommand{\p}{\partial}

\newcommand{\be}{\begin{equation}}
\newcommand{\ee}{\end{equation}}
\newcommand{\la}{\label}

\newcommand{\rt}{\rm}

\begin{document}

\title[Regularizations of Laplacian Growth]{Integrability-preserving regularizations of Laplacian Growth}

\date{December 2018}

\author[R. Teodorescu]{Razvan Teodorescu}
\email{razvan@usf.edu}
\address{4202 E. Fowler Ave., CMC342, Tampa, FL 33620}

\keywords{integrable systems, free boundary problem, quadratic differentials}
\subjclass{Primary: 30D05, Secondary: 30E10, 30E25}

\begin{abstract}
The Laplacian Growth (LG) model is known as a universality class of scale-free aggregation models in two dimensions, characterized by classical integrability and featuring finite-time boundary singularity formation. A discrete counterpart, Diffusion-Limited Aggregation (or DLA), has a similar local growth law, but significantly different global behavior. For both LG and DLA, a proper description for the scaling properties of long-time solutions is not available yet. In this note, we outline a possible approach towards finding the correct theory yielding a regularized LG and its relation to DLA. 
\end{abstract}

\maketitle

\section{Introduction}

There is a large number of  models for physical processes involving moving boundaries in two dimensions which can
be reduced to the same universality class, known either as {Hele-Shaw processes} or {Laplacian Growth} (LG). Examples of such physical phenomena include: electrodeposition \cite{Gollub}, solidification \cite{Langer},  viscous
fingering \cite{LG-review}, bacterial and cancer cells growth  \cite{BenJacob},  to illustrate only a few of the experimental areas where this model is widely used. The mathematical formulation of this dynamical system can be traced back many decades ago, to the pioneering work of  Polubarinova-Kochina and Galin \cite{Galin, Polubarinova-Kochina} and has since generated a lot of interest, owing in particular to its applicability to a large number of different research fields (see \cite{GTV} and references therein), such as: propagation of crystallization fronts, dielectric breakdown and structural failure (shock) propagation, the stochastic L\"owner equation, inverse spectral problems for non-normal operators, integrable hierarchies, shape recognition and domain reconstruction in computer tomography, and the list could continue.

Considered from a fluid dynamical point of view, the process can be described as follows. 
In two dimensions, two incompressible and immiscible fluids (one viscous and the other inviscid) are separated by an infinitely-thin boundary (modeled by a real-analytic Jordan curve). The dynamics of the boundary is governed by Darcy's law, according to which the normal component of the boundary velocity is proportional to the normal component of the pressure gradient.  If, due to external factors (or to initial conditions), the interface develops a region of high curvature (a ``bump") at some point in time, the pressure gradient on that bump will be larger than on the rest of the boundary. The bump therefore will be moving (``growing") faster than the flatter portions of the boundary, and will be quickly amplified during the growth process. The growth process is thus
very unstable and may produce asymptotic shapes such as long, thin ``fingers'' \cite{DiBenedetto, Taylor} or, in the long time limit, a fractal-like cluster with characteristics similar to those obtained in the Diffusion Limited Aggregation (DLA) processes, which is a diffusion-governed variant of the same growth law, cf., e.g.,  \cite{BenJacob, Praud-Swinney, Gollub}. 

The mathematical model governing these processes reduces to the following equation for the dynamics of the 
moving boundary: 
\begin{equation}\label{growthEq}
V(\zeta) = \partial_n G_{D(t)}(\zeta,a).
\end{equation}
Here $V$ is the normal component of the velocity of the moving
boundary $\partial D(t)$ of the time-dependent domain $D(t)
\subset {\mathbb C}$, $\zeta \in
\partial D(t)$, $t$ is time, $\partial_n = \widehat{N}\cdot \vec{\nabla}_{\zeta}$ is the normal component of
the gradient, and $G_{D(t)}(\xi,a)$ is the Dirichlet Green's  function of the
domain $D(t)$ for the Laplace operator with a unit source located at
the point $a \in D(t)$. Equation (\ref{growthEq}) is known as the {Darcy law} (in recognition of the original, empirical 
law derived by Darcy in describing the effective flow of water through porous solids, under the influence of gravity, c.f. \cite{GTV} for details).
 
In DLA, a random walker is released from $a=\infty$, and wanders into $D(t \in \mathbb{N})$ with absorbing boundary conditions, with probability of first-time exit through a subset $\gamma$ of the boundary $\p D(t)$ given by the harmonic measure $d \omega(\gamma, a)$. The discrete time $t$ is then incremented by one unit, and the process is repeated.

The contribution of Polubarinova-Kochina and Galin to the theory led to the reformulation of this problem as the area-preserving diffeomorphism identity
\begin{equation}\label{areapres}
\Im \,(\bar f_t f_{\theta}) = 1,
\end{equation}
where $f(t, . )(\mathbb{T}) = \partial D(t)$ is the moving boundary (assumed smooth), 
parameterized by $\theta \in \mathbb{T} = \p \mathbb{D}$, and conformal when
analytically extended into the region  $\Im \theta 
\geq 0$ \cite{Galin, Polubarinova-Kochina}.  The equation
(\ref{areapres}) is associated to remarkable properties, such as the existence of a countably-infinite set of
conservation laws:
\begin{equation}\label{moments}
C_n(t) = \frac{1}{\pi}\int_{D(t)}\, z^n \, dx\,dy = C_n(0),
\end{equation}
where $n \in \mathbb{Z}$ is $n \ge 0$ \cite{Richardson1972}
(or $n \le 0$ \cite{Mineev1990}) in the case of an finite
(infinite) domain $D(t)$, and exact
time-dependent closed-form solutions \cite{Varchenko}. 

Unfortunately, as shown in \cite{GTV} and many other works, this dynamical 
process is known to develop boundary singularities in finite time, which do 
not admit a continuation in the space of strong (classical) solutions. While the 
problem of existence of an infinite-time strong solution (except for the simple 
geometry corresponding to homothetic ellipsoidal domains \cite{Varchenko}) is open, there is 
a large class of boundaries for which evolution towards a singularity in finite time 
has been established \cite{GTV}. 

This paper is organized as follows: in the next section, a summary review of 
recent attempts at regularizing this dynamical process with infinite-dimensional 
symmetry group (both for LG and DLA) is provided. This survey discusses both 
the mathematical component and the physical interpretation of the works reviewed. 
In Section 3, a new proposal for a possible solution is introduced. Based on the weak 
formulation of LG, this approach has the advantage that a solution is available for all 
times, while the physical interpretation (in the sense of quantum deformations) relates 
the new formulation to the concept of quantum phase transitions. A brief comments section 
concludes this paper. 

The author wishes to thank the hospitality of the Mathematical Institute at K\"oln University, where 
this work was initiated in May 2018. 

\newpage

\section{Integrability-preserving regularization: a brief history}

There have been several different attempts at identifying a more general 
dynamical system which would naturally incorporate LG and DLA as 
special limits, retaining the integrability structure of LG while remaining 
compatible with the rescaling invariance of DLA. The difficulty can be 
easily uncovered, from each of these two points of view: as an integrable 
system, LG is equivalent to the dispersionless limit of the Toda hierarchy \cite{2DToda}, 
and therefore does not have any special lenthgscale, which is another 
way of saying that its integrable structure is conformal in nature. In order 
to provide a (short-scale) regularization mechanism, it would appear that 
this symmetry needs to be broken, which would do away with all the features 
that make LG an interesting process to begin with. 

On the other hand, DLA is in turn marked by its formulation as a discrete 
aggregation process, and by the difficulty of defining probability measures 
on spaces of curves. When attempting to describe DLA by starting from 
purely discrete probability measures (such as in the recent works \cite{AW1, AW2, AW3}), 
the main difficulty stems from the fact that do not have yet a generalized 
Central Limit Theorem for aggregation processes of these types. 

The need for finding an extension to both LG and DLA is due to the inherent 
limitations of both of these models, when trying to use them to describe 
complex growth patterns occurring in nature. For example, it is known from 
experiments that 2D growth processes governed by the boundary Darcy law 
do not exhibit finite-time singularities, but instead the process continues by 
means of viscous fingering, which converts a high-curvature local instability 
into a branching process \cite{Praud-Swinney}. However, classical solutions 
for LG with the same type of initial conditions cannot be continued past a 
boundary singularity formation \cite{Howison86}. 

Likewise, while computer simulations of DLA can successfully produce approximations 
to various fractal dimensions of the DLA cluster, a rigorous theory for this process 
is not available yet, despite several attempts at introducing a proper short-scale regularization 
by a number of mechanisms \cite{CM01, JST12, JST15, NT12}. 

We briefly survey the recent attempts at describing either (or both) LG and DLA 
as special limits of an integrable dynamical process with a short length-scale regularization, 
after which we provide an alternative approach. 

\subsection{Normal random matrix ensembles}

In \cite{TBAWZ05} and subsequent publications, LG was linked to the equilibrium 
distribution of an ensemble of normal random matrices, in a prescribed external field.  
The {{Normal Random Matrix Ensemble}} is defined through the statistical weight associated with a function  $V(M,M^{\dag})$, where $[M, M^{\dag}]=0$, so that both $M$ and $M^{\dag}$
can be diagonalized simultaneously: 
\be
\label{ZN}
d \mathbb{P}(M) = e^{-N{\rm tr} \, V(M, M^{\dag})} d\mu (M).
\ee
Here $N \in \mathbb{N}, N \gg 1$ is a parameter describing the size of matrices considered,   and the measure of integration
over normal matrices
is induced by the flat metric
on the space
of all complex matrices ${d}_C M$, where
${d}_C M = \prod_{ij}d\, {\rm Re} \, M_{ij} d\, {\rm Im} \, M_{ij}$.
Using a standard procedure,
one passes to the joint
probability distribution
of eigenvalues $z_1,\dots,z_N$:
\be
\label{mean}
d \mathbb{P}(\{z_1,\dots,z_N\})=  \frac{1}{ \tau_N}|\Delta_N (z)|^2 \,\prod_{j=1}^N
e^{-N V(z_j,\bar z_j)} {d}^2 z_j
\ee
Here
$d^2 z_j \equiv d x_j \, d y_j$ for $z_j =x_j +iy_j$,
$\Delta_N(z)=\det (z_{j}^{i-1})_{1\leq i,j\leq N}=
\prod_{i>j}^{N}(z_i -z_j)$
is the Vandermonde determinant, and
\be
\label{tau}
\tau_N = \int
|\Delta_N (z)|^2 \,\prod_{j=1}^{N}e^{-N
V(z_j,\bar z_j)} d^2z_j
\ee
is a normalization factor, the partition function
of the matrix model (also called tau ($\tau$)-function). 
It is in fact the tau-function $\tau_{N}(t)$ of the 2D Toda hierarchy, whose large $N$ limit 
describes the solution of the Laplacian growth equations \cite{WZ}, 
\begin{equation}\label{tau1}
\tau_{N}(t)=\idotsint\limits_{{\Bbb C}^{N}}\prod_{i\neq j}|z_{i}-z_{j}|^{2}e^{-N\sum_{k=1}^{N}V(z_{k})}d^{2}z_{1}\cdots d^{2}z_{N},
\end{equation}
in the case when the function $V(z)$ has the form
\begin{equation}\label{tau2}
V(z) = \frac{1}{t_0}\Bigl(|z|^2 - Q(z) -\overline{Q(z)}\Bigr),\quad 
Q(z) = \sum_{k=2}^{d+1}t_kz^k,\quad t_k \in {\Bbb R}.
\end{equation}
The parameter $t>0$ is a time variable.
 At the same time,  the right hand side of (\ref{tau1}) represents the
partition function of the distribution of eigenvalues in a {\it normal matrix model}, hence  the key
problem is to study the large $N$ limit of the multiple integral in the right hand
side of equation (\ref{tau1}). The eigenvalue density is given by

$$
\rho_N(z) \equiv \frac{1}{\tau_N} \int \sum_{j=1}^N \delta(z-z_j)
|\Delta_N (z)|^2 \,\prod_{j=1}^{N}e^{-N
V(z_j,\bar z_j)} d^2z_j
$$

Assume that we are given a domain $D$ of normalized area $t$ and harmonic moments \eqref{moments} $t_k = \frac{C_k}{k}, k \ge 1$, and the function $V(z)$ 
satisfies the constraint
\be \la{qwe}
\int_{\mathbb{C}} |z|^n e^{-N V(z)} d^2 z < \infty \quad n =0,1,2,\dots
\ee
for all for values of the scaling parameter $N>0$.
For fixed $N \in \mathbb{N}$, the orthogonal polynomials $\{ P^{(N)}_n(z) \}$ with weight function $e^{-NV(z)}$ are 
defined by
\be
\int_{\mathbb{C}} P^{(N)}_n(z) \overline{P^{(N)}_m(z)} e^{-NV(z)} d^2 z = 
\delta_{nm}. 
\ee
The approximation method presented in this approach is based on the following 
statement: as 
$n \to \infty, \,\, N \to \infty, \,\, n \to Nt, \,\, n \le N,$
the squared norms of weighted polynomials $|P^{(N)}_n(z)|^2 e^{-NV(z)}$ (which we denote by $\mu_n^{(N)}(z)$ 
in the following) converge to the harmonic measure of the domain $D(t)$, 
with support  $\Gamma(t) = \p D(t)$:
\be \la{limit}
\mu_n^{(N)}(z) = 
|( f^{-1}(z))'|  \left  [ 1+ O \left ( \frac{1}{N} \right ) \right ],
\,\, n, N \to \infty,
\ee
where $f(z)$ maps conformally the exterior of the unit disc to the exterior of the domain $D(t)$. 

While this approach does indeed provide a properly regularized, integrability-preserving deformation of LG, it has not produced so far solutions that could explain the mechanism of viscous fingering, with the notable exception of the special compressibility shock solutions of  \cite{Lee-Teodorescu-Wiegmann, Lee-Teodorescu-Wiegmann1, Lee-Teodorescu-Wiegmann-new}, for which a different physical interpretation is available. However, as indicated in Section 3, this approach does lead to a different class of solutions, at the cost of abandoning the original boundary dynamics as the primary formulation of the dynamical process. 

\subsection{Discretized ``path integrals" and electrostatic clusters: DLA models}

\subsubsection{DLA-like models as aggregation of electrostatic clusters}

Consider the dynamical process given by the time-dependent family of conformal maps sending $\mathbb{D}$ into $\Omega(t)$, and satisfying 

\begin{equation}
\frac{\partial f(z,t)}{\p t} =z f'(z,t)\frac{1}{2\pi}\int_{\mathbb{T}}\frac{z+\zeta}{z-\zeta}\frac{|d\zeta|}{|f'(\zeta,t)|^{\alpha}}, 
\label{bensimonshraiman}
\end{equation}
where $\alpha\in \mathbb{R}$ is a model parameter. As shown in \cite{LG-review, CM01}, \eqref{bensimonshraiman} can be related to  the Dirichlet Green's functions $G(z,t)$ of the domains $\Omega(t)$, with a source at $\infty$: if $V_n$ denotes the normal velocity of the boundary, then $V_n\sim |\nabla G(\cdot, t)|^{\alpha-1}$, therefore describing a more general class of LG-like processes known as Dielectric Breakdown Models (DBM). 

For $\alpha=2$ in \eqref{bensimonshraiman}, we retrieve a formulation of Laplacian Growth, as shown, e.g.,  in  \cite{GTV} and references therein. The limiting case $\alpha=0$ reduces  (after applying the Poisson integral formula with constant Dirichlet data) \eqref{bensimonshraiman} to 
\[\partial_t f(z,t)=zf'(z,t)\]
for which a class of solutions is given by $f(z,t)=\varphi(e^tz)$, for any conformal map $\varphi$. As the local growth law has velocity inversely proportional to the curvature, any initial real-analytic boundary will be smoothed out further by the process, with the curvature converging to a constant. In particular, the solution $f(z,t)$ exists for all $t>0$, and the infinite-time limiting shape is a circle.

As discussed in \cite{JST15} (from where this argument is borrowed), there is another case where \eqref{bensimonshraiman} can be solved for all $t>0$: setting an arbitrary $\alpha\in \mathbb{R}$, and the initial condition $\varphi(z)=z$, then 
\begin{equation}
f(z,t)=(1+\alpha t)^{\frac{1}{\alpha}}z
\label{limitmap}
\end{equation}
solves the initial-value problem, and the solution is again smooth, for all $t>0$ when $\alpha\geq 0$, and for $t<-1/\alpha$ for negative $\alpha$. 

The conformal maps in \eqref{limitmap} correspond to the small-particle scaling limit for the Hastings-Levitov model with strong regularization \cite{Hastings-Levitov}. The authors conjectured that the growth behavior in the Hastings-Levitov random aggregation model will exhibit a phase transition at the parameter $\alpha=1$, indicating the separation between boundary dynamics dominated by smoothing (for $\alpha < 1$) and curvature-driven instability (for $\alpha > 1$). 

Starting from this formulation for the DBM class of dynamics, various ad-hoc regularizations were introduced, e.g. by adding a small parameter $\sigma > 0$ to $f'(t, z)$ under the integral sign in \eqref{bensimonshraiman}, which basically shifts the boundary curves away from the level lines of the Green's function (for the simplest case $\alpha = 0$), therefore preventing the velocity (left-hand side of \eqref{bensimonshraiman}) from diverging. However, the selection and even asymptotic limit ($\sigma \to 0$) of the new parameter seem to be important for the type of regularized solutions obtained, which is indicative of the singular-perturbation character of LG. 

The main open problem that is not addressed by this formulation is the branching behavior observed in numerous computer simulations of DLA. The theory does not seem to have any mechanism for this type of instability (which corresponds to the viscous fingering of LG), and therefore cannot provide a method for computing the fractal dimensions of large DLA clusters. 

\subsubsection{Probabilistic models for DLA and asymptotic theory}

In  \cite{AW1, AW2, AW3}, an alternative approach to defining DLA was provided, based on the expected weak convergence of probability functions describing a discretized boundary, and starting from a staking process for identical objects of given size, which is then rescaled in a prescribed manner. We reproduce here the main idea of this approach, referring the reader to the works cited for more details. 

To define the growth process,  the unit circle is partitioned into $N \gg 1$ equal segments of length
$\Delta \phi = \frac{2\pi}{N}$. Their images on the boundary of the cluster, under the conformal map $f(z, t)$, will have lengths
\begin{align}
l_k= |f'(z_k, t)| \Delta \phi = \frac{2\pi}{N} |f'(z_k, t)|,
\label{block length}
\end{align}
where $z_k = e^{i \phi_k}$, where $\phi_k \in [k\Delta \phi, (k+1)\Delta \phi]$. The process consists of rectangular ``blocks", each landing on the segments $\gamma_i = f(\, ., t)([i\Delta \phi, (i+1)\Delta \phi]) \subset \Gamma(t), \, i = 1, 2, \ldots, N$. A block landing on the i-th segment is prescribed the ``height"
\begin{align}
h_i = \frac{\epsilon}{l_i},
\label{block height}
\end{align}
where $0 < \epsilon \ll 1$ is a regularization parameter, thus ensuring equal areas.

Assume now that during a single time step $k_i \geqslant 0$
blocks land on the segment $i$. Then the growth step is fully
specified by the set of $N$ non-negative integers $k_i$ (occupation numbers).  For a given total number of blocks, $K$, the occupation numbers satisfy the constraint
\begin{align}
\sum_{i=1}^N k_i = K.
\label{constraint-discrete}
\end{align}
If all the building blocks attach to the cluster independently,
the probability of a particular growth step $\{k_i\}_{i = 1}^N$ is given
by the multinomial distribution, 
\begin{align}
P(\{k_i\}) = \frac{K!}{N^K \prod_{i=1}^N k_i!}.
\label{probability-discrete}
\end{align}

According to the authors,  the displacement
$\delta n_i$ of the segment $\gamma_i$  in one
time step is computed as:
\begin{align}
\delta n_i &= k_i h_i = \epsilon \frac{k_i}{l_i} =
\frac{\epsilon N}{2\pi} \frac{k_i}{|f_t'(z_i)|} 
\end{align}
Identifying the displacement (per unit of discrete time) to 
velocity, the corresponding Darcy-type growth law becomes 
\begin{align}
k_i = \frac{2\pi}{\epsilon N} |f_t'(z_i)| v_i 
\end{align}
where $v_i = \delta n_i $ is the normal ``velocity" of the boundary.

The family of models is parameterized by $\kappa := \frac{K}{N}$, 
understood as the average number of building blocks deposited to a single segment on the
boundary. It is claimed that the limit $\kappa \ll 1$ corresponds to DLA.

Independently, a second model parameter is the ``area" of a building block, 
$\epsilon$, with the limit case $\epsilon \to 0$ yielding LG. 

The discrete-time process described above is related to a continuous-time 
stochastic process $X_t(\phi)$ by taking the limit  (in distribution) of the random variable $\frac{k_i}{\kappa}$, as $N \to \infty, \, K \to \infty,$ at a fixed value of $\kappa, \, \epsilon$. 

After computing a Freidlin-Wentzell type of functional for large deviations of the process $X_t$ from its mean $\mathbb{E}(X_t) = 1$, various limits are considered, such as $\epsilon \to 0$ (supposed to lead to a deterministic process equivalent to LG) and $\kappa \to \infty$ (for which the large deviations functional becomes the Cram\'er functional for a Poisson process).  It is expected that an application of Varadhan's principle for this functional will, for the proper choice of parameters $\epsilon, \, \kappa,$ yield DLA as the solution to a variational problem. 

\subsection{Quantum deformations and representation theory for CFT}

In \cite{Eldad}, a symmetry-based approached was discussed as a potential way of embedding LG (and, supposedly, DLA) into a wider integrable system, whose representations would incorporate finite-dimensional reductions corresponding to those of rational conformal field theories (RCFT), as described in many works, see, e.g., \cite[\S 7-8]{GTV}. The 
justification for this approach lies in the identification of  LG with the dispersionless Toda hierarchy, the Gelfand-Dickey hierarchy, and the KP hierarchy \cite{GTV, 2DToda,TBAWZ05, WZ}. As the latter is known to contain the Virasoro algebra (at arbitrary values of the central charge), the connection with RCFT is to be expected. 

However, such an embedding has not been established rigorously, yet. The main obstacle for this approach stems from the limitations of quantum deformations, as discussed briefly in the following. On one hand, the most general procedure for constructing quantum deformations is the Drinfeld method \cite[ \S 6]{QGroups}, which yields a quantum double (pair of Hopf algebras) for every compact Lie group (classical symmetry). In this approach, an (arbitrary) deformation parameter is naturally introduced, which should appear as the desired regularization parameter needed in order to resolve the singularities emerging in the classical (undeformed) problem. The limitation is due to the fact that the method is restricted to classical symmetries which correspond to finite-dimensional Lie groups, unlike the case of LG. 
 
In those situations where LG dynamics can be mapped, locally, to a finite-dimensional reduction of an integrable hierarchy, the procedure works as expected. For example, boundary singularities corresponding to reductions to an elliptic curve (known as the $(3, 2)$-cusps) were considered in \cite{Lee-Teodorescu-Wiegmann-new} and \cite{TWZ05}, leading to the KdV and Painlev\'e I equations, respectively. This is an illustration of the well-known fact that integrable deformations of the field of elliptic functions lead to the Painlev\'e I transcendent \cite{Lee-Teodorescu-Wiegmann-new}, which is also a scaling reduction of the KdV equation \cite{TWZ05}. 

Another classical-quantum relationship was considered in \cite{Eldad}, based on the observation that (again, in the case of quantum deformations corresponding to an elliptic curve) certain classical Poisson structures are naturally embedded into a  Sklyanin algebra, depending on an auxiliary parameter, i.e. the modulus of the elliptic curve. 

The difficulty in applying such a program to LG in general is due to the incomplete classification of boundary singularities arising from the classical dynamics. In the absence of a probabilistic description of various types of possible boundary singularities (e.g., it is known that classical solutions cannot be continued in the cases of cusps of types $(4k-1, 2), \, k \in \mathbb{N}$, \cite{GTV}, but the likelihood of any such cusp formation has not been quantified), it is not possible to apply the deformation method 
in order to compute fractal dimensions and critical exponents.      

In the remaining of the paper, we investigate another approach towards an integrable deformation of LG, and its application to the case of boundary singularity formation.  

\newpage 

\section{Weak solutions and quantization}

\subsection{Equilibrium measures and orthogonal polynomials}

\subsubsection{The classical equilibrium problem for logarithmic potentials on $\mathbb{C}$}

Let $\sigma$ be a finite Borel measure  on the plane, and define its associated {logarithmic potential}
$
 \Phi^{\sigma}(z)\equiv  -\int \log |z-x|\,d\sigma (x),
$
{logarithmic energy} 
$
W[\sigma] \equiv -\iint \log |z-x|\,d\sigma (x)d\sigma (z),
$
both not necessarily finite. Denoting by $\supp (\sigma)$ the support  of $\sigma$, consider an admissible real-valued function $V$, called {external field}. Accordingly, we obtain the {total potential}
\begin{equation} \label{totalPotential}
U^{\sigma, V}(z) \equiv  \Phi^{\sigma}(z)+V(z),
\end{equation}	
and the {total energy}
\begin{equation} \label{totalcontEnergy}
\mathcal{E} [\sigma, V]= W[\sigma] + 2\int V(z) \, d\sigma(z).
\end{equation}	

Given a class of measures (e.g., atomic, absolutely continuous, defined on a compact, etc.), we define the equilibrium measure as the minimizer, within that class, of the total energy  $\mathcal{E} [\sigma, V]$ (including for the case $V = 0$, in which case the equilibrium measure is a characteristic of the set on which it is supported). 

For example, given a compact set $K \subset \mathbb{C}$, its equilibrium measure (when it exists) is known as the {Robin measure} of $K$. Denoted $\omega_K$, this measure is unique, supported on the outer boundary of $K$, and its energy is given by 
\begin{equation} \label{robin}	
 W[\omega_K] = - \ln \cp \, K,
\end{equation}
where $\cp \, K$ is the logarithmic capacity of $K$ (discussed in some detail in the next subsection). However, the equilibrium measure in a (non-trivial) external field is a much more complicated object. In particular, it is not know a priori where its support will be located.

\subsubsection{Equilibrium measures for random matrix ensembles}

Let $t>0$ and choose a closed subset $\Sigma\subset \mathbb{R}$. If we denote by $\MM_t(\Sigma)\,$ the set of positive Borel measures $\sigma$ compactly supported in $\Sigma$ with total mass $t\,,$ for a real-analytic external field $V(x)$ on $\mathbb{R}$, the subset of measures $\sigma \in  \MM_t$ with finite energy is not empty, see \cite[Ch.~I]{Saff:97}). Therefore, there exists a unique minimizer measure, denoted 
$\mu_t(V)\in \MM_t$,  such that
\begin{equation}\label{weightedenergy}
\mathcal{E}[\mu_t, V]= \min_{\sigma \in \MM_t } \mathcal{E}[\sigma, V].
\end{equation}
If the function $V(x)$ (sometimes called ``confining potential" due to its physical interpretations) grows at infinity faster than $\log |x|$, then  $\supp(\mu_t)$ is a  compact set in $\mathbb{R}$ for all $t > 0$, and $\mu_t$ is completely determined by the equilibrium condition satisfied by the total potential 
\begin{equation}\label{equilibrium1}
U^{\mu_t, V}(z) = u_{t}, \, \forall \,  z \in {\supp} (\mu_t), \,  U^{\mu_t, V}(z) \geq u_{t}\, \forall \,    z \in \mathbb{R}, 
\end{equation}
where the constant $u_t$ is given by 
\begin{equation} \label{extremalConst}
u_{ t} := \frac{1}{t}\left( \mathcal{E}[\mu_t, V]-\int V(x) \, d\mu_t(x) \right).
\end{equation}	

An application of the notion of equilibrium measures for orthogonal polynomials and random matrix models \cite{BKD, BI, BK1,  BK} is based on the following fundamental result from \cite{Gonchar:84, Mhaskar/Saff:85} (see also \cite{MR916090} and \cite{Saff:97}): for a fixed external field $V$ and $N > n > 0$, let $(. \, , .)$ be the scalar product on $C(\mathbb{R})$, with measure $e^{-2N V(x)} dx$, and induced norm $|| . \, ||$, and define the orthogonal monic polynomials $Q^{(N)}_{n}$, such that $(Q^{(N)}_n, x^k) = 0, \quad k=0, 1, \dots, n-1$. These polynomials are associated with $n \times n$ Jacobi matrices, as a representation of their recurrence relations. For $n, N\to \infty$ at fixed $\frac{n}{N} = t >0$,  the  zero-counting measure  
$
\chi_n := \frac{1}{n} \mathbb{I}_{\{ Q^{(N)}_n(x) = 0 \}}
$
converges (in the weak star topology) to the equilibrium measure $\mu_t(V)\in \MM_t$, which is also  the limit spectrum of the infinite Jacobi matrix, associated with the scaling $\frac{n}{N}\to t$. Moreover,  $\frac{\log ||Q^{(N)}_n||}{N}$ has the non-zero limit  $u_t$, the equilibrium potential constant. 

The connection to random matrix ensembles arises when considering the class of uniform atomic distributions. Denote by $M_n$ the class of such distributions supported on $n$ points on the real line $\bm \zeta=(\zeta_1, \dots, \zeta_n)\in \R^n$,   which can be identified to the atomic distribution itself, $\bm \zeta\in \R^n=\nu_n \in M_n$, then the corresponding discrete energies are defined by (cf.\ \eqref{totalcontEnergy})
\begin{equation} \label{nn}
E[\bm{\zeta}, V] := E[\nu_n, V]:=   \sum_{i\neq j} \log\frac{1}{|\zeta_i-\zeta_j|} + 2 \int V(\zeta) d\nu_n(\zeta)
\end{equation}
Introducing the joint atomic probability mass density 
$$
d \mathbb{P}^{(N)}_{n}(\bm \zeta) = \frac{1}{Z^{(N)}_{n}}\, e^{-E[\bm{\zeta}, N V]} d\bm \zeta , \quad  Z^{(N)}_{n}=\int_{\R^n } e^{-E[\bm{\zeta}, N V]} d\bm \zeta, 
$$
we first notice that it is identical to the probability density for eigenvalue distribution of the normal random matrix model, when the spectrum is restricted to the real line (which, in fact, describes the so-called the Gaussian Unitary Ensemble, perturbed by the potential function $N\cdot V(x)$). Then the  classical Heine formula
 $
 Q^{(N)}_n(x)=\int_{\R^n} \prod_{k=1}^n (x-\zeta_k)d \mathbb{P}^{(N)}_{n}(\bm \zeta) 
 $
shows that the orthogonal polynomial $Q^{(N)}_n$ 
is the average of polynomials with real zeros randomly distributed according to the probability density $d\mathbb P^{(N)}_{n}$.

\subsection{Minimal capacity sets in external fields and moduli functions}

We review the main results characterizing minimal capacity sets in $\mathbb{C}$, and how they relate to the problem of finding the support of equilibrium measures, following \cite{GTV}. Let $\Omega$ be a domain in $\mathbb C$ and $\rho (z)$ be a
real-valued, Borel measurable,  non-negative function in $L^2(\Omega)$, defining a differential metric $\rho$ on $\Omega$ by $\rho:=\rho(z)|dz|$. For any locally rectifiable curve $\gamma \subset \Omega$, the 
integral $l_{\rho}(\gamma):=\int_{\gamma}\rho (z) |dz|$, is called the $\rho$ -- length of $\gamma$, while the $\rho$ -- area of $\Omega$ is given by 
\begin{equation}
A_{\rho}(\Omega):=\iint\limits_{\Omega}\rho^2 (z) dx dy.
\label{eq:1.2}
\end{equation}

If $\Gamma$ is a family of curves $\gamma$ in $\Omega$, we define the modulus of $\Gamma$ in $\Omega$ by 
$$
m(\Omega,\Gamma)=\inf\limits_{\rho}\frac{A_{\rho}(\Omega)}{L^2_{\rho}(\Gamma)}, \,\, 
L_{\rho}(\Gamma):=\inf\limits_{\gamma\in\Gamma}l_{\rho}(\gamma),
$$
where $L_{\rho}(\Gamma)$ is the $\rho$-length of the family $\Gamma,$ and the infimum is taken over all metrics $\rho$ in $\Omega$.

%

The  modulus is conformally-invariant, and monotonic relative to inclusion of curve families: let $\Gamma$ be
 a family of curves in a domain $\Omega\in \widehat{\mathbb C}$,
and let $w=f(z)$ be a conformal map of $\Omega$ onto
$\widetilde{\Omega}\in \widehat{\mathbb C}$. If
$\widetilde{\Gamma}:=f(\Gamma)$, then
$m(\Omega,\Gamma)=m(\widetilde{\Omega},\widetilde{\Gamma}).$
Also, if $\Gamma_1\subset\Gamma_2 \subset \Omega$, then
$m(\Omega,\Gamma_1)\leq m(\Omega,\Gamma_2)$. 

\subsubsection{Capacity and reduced modulus} \label{sec}

 Let $\Omega\subset \widehat{\mathbb C}$ be a simply connected domain, and $a\in \Omega$,
$a \ne \infty$. For the doubly connected domain
$\Omega_{\epsilon}=\Omega\setminus B_a(\epsilon)$, $0 < \epsilon \ll 1$, the reduced modulus of $\Omega$ with respect to the point $a$ is defined as 
$$
M(\Omega,a):=\lim\limits_{\epsilon\to
0}\left(M(\Omega_{\epsilon})+\frac{1}{2\pi}\log\,\epsilon\right), 
$$  
where $M(\Omega_{\epsilon})$ is the modulus of the domain $\Omega_{\epsilon}$ with respect to the family of curves that separate its boundary components. As shown in \cite{Jen3}, it is also equal to $\frac{1}{2\pi}\log\,R(\Omega,a)$, where $R(\Omega,a)$ is the conformal radius of  $\Omega$ with respect to $a$.

The reduced modulus $M(\Omega,\infty)$ of a simply
connected domain $\Omega$,  with respect to $\infty\in \Omega$ is defined as the reduced modulus of the image of $\Omega$ under the map $1/z$, 
$$
M(\Omega,\infty)=-\frac{1}{2\pi}\log R(\Omega,\infty). 
$$ 


 For two disjoint compact sets $K_1$, $K_2$ in $\mathbb{C}$, the capacity of the 
 complement (often called {\it{condenser}}, owing to the connections to electrostatics) is defined as $\emk (\widehat{\mathbb{C} }\setminus (K_1 \cup K_2) )
 := \inf \mathcal{E}_D (\widehat {\mathbb{C}} \setminus (K_1 \cup K_2))$, where $\mathcal{E}_D$ stands for 
 Dirichlet energy, and the infimum is taken over all Lipshitz-continous functions on the set. 
 
A condenser is said to be {admissible} if there exists a solution to 
the Dirichlet problem on $\widehat{\mathbb{C} }\setminus (K_1 \cup K_2)$, 
continuous on $\widehat{\mathbb{C}}$, taking the value 0 on $K_1$ and 1 on $K_2$. 
Therefore, for admissible condensers, the capacity is a conformal invariant. This can be used to compute it for the case 
when  $K_1$ and $K_2$ are two disjoint continua (compact and connected), yielding $\emk\, C=2\pi/\log R$, where 
$R > 1$ corresponds to the conformal mapping of $\widehat{\mathbb C}\setminus \{K_1\cup K_2\}$ onto an
annulus $1<|w|<R$. 


Let now $K$ be a compact  set in $\mathbb C$, and condensers
of special type $C_R=\{|z|\geq R, K\}$ for $R \to \infty$. The function
$\frac{1}{\emk\,C_R}-\frac{1}{2\pi}\log\,R$ increases with
increasing $R$ and the limit
\begin{equation}
\emk\,K=\lim\limits_{R\to\infty}{R}\exp\left(-\frac{2\pi}{\emk\,C_R}\right)\label{eq:1.6}
\end{equation}
exists and is said to be the {\it logarithmic capacity} of the
compact set $K\subset \mathbb C$. The result (\ref{eq:1.6}) is 
known as Pfluger's theorem, \cite[\S 9]{Pom2}.
.

Next we briefly summarize the definition and some properties of
the logarithmic capacity of a compact  set $K\subset \mathbb C$
following Fekete. For $n=2,3,\dots$ we consider $$
\Delta_n(K)=\max\limits_{z_1,\dots,z_n\in K}\prod\limits_{1\leq
k<j\leq n}^n|z_k-z_j|. $$ The maximum exists and is attained for
so-called Fekete points\index{Fekete points} $z^{(n)}_{k}\in \partial
K$, $k=1,\dots, n$. Its value is equal to the
Vandermonde determinant
$$ \Delta_n(K)=\Big|\det\limits_{j, k=1,\dots,
n}(z^{(n)}_{k})^{j-1}\Big|. $$ Then, the limit 
\begin{equation} \label{new}
\lim\limits_{n\to\infty}(\Delta_n(K))^{\frac{2}{n(n-1)}} = \emk\,K 
\end{equation}
exists (see \cite{Pom2}), is known as the {transfinite
diameter} of $K$, and is equal to its logarithmic capacity. 

This result establishes the connection 
between the logarithmic capacity of a compact set (support of the equilibrium measure) 
\eqref{robin} and the $n \to \infty$ limit of random matrix ensembles, as discussed in the 
previous section.


\subsection{Weak solutions of LG and equilibrium measures}

Given a solution to LG for some initial data, in the form of a family of domains $\{\Omega(t) \}_{t \in [t_0, T]}$ with the strict inclusion property $\Omega(s) \subset \Omega(t),\, \forall t_0 \le s < t \le T$, consider its associated Riemann surface \cite[\S 5]{Mineev-Put-Teo},  and its set of branch points, denoted by ${\bm {B}}(t)$.  For this data, we then solve the (Chebotarev) problem of finding a set of analytic arcs with endpoints in $\bm{B}(t)$, $K(t)$, and the equilibrium measure supported it, solving \eqref{weightedenergy}. 

Denote the equilibrium measure by $\mu(t)$ (functionally dependent on $V$), then we refer to the family of sets and measures $\{(K(t), \mu(t) \}_{t \in [t_0, T]}$ as the weak solution to LG with prescribed branch point data. It should be noted that, owing to the relation between this problem and the inverse balayage of the domains $\{\Omega(t)\}$, we have 
$K(s) \subseteq K(t),\, {{\supp} \, \mu(s)} \subseteq {{\supp} \, \mu(t)},\, \forall t_0 \le s < t \le T$, c.f. \cite{Mineev-Put-Teo}. 

Even in the (typical) case when the (strong) LG solution develops cusp boundary singularities for some value of the area, $t_c$, therefore forbidding continuation of the strong solution for $t \ge t_c$, the weak solution remains well-defined, but the support of the equilibrium measure branches out at the transition through $t=t_c$ \cite{BK, Lee-Teodorescu-Wiegmann-new}.  

To describe both the weak solution in potential-theoretical sense and the integrable character of LG, it is useful to define the function 
$$
\Psi(z, t; V) := \exp \left ( V(z) + \int \log|z - \zeta| d\mu_t(\zeta) \right ), \,\, z \notin {\supp} \, \mu(t).
$$
It is subharmonic in a neighborhood of $\supp \, (\mu)$ (excluding the support), and in a neighborhood of  $\infty$, and superharmonic in a tubular neighborhood of $\p \Omega(t)$.  Since, up to a prefactor that is independent on $t$, the function is proportional to the ratio 
$$
\exp (\mathcal{E}[\mu_t, V(\zeta) + \log|z-\zeta|] / \exp (\mathcal{E}[\mu_t, V(\zeta)],
$$
that is the limit (in capacity) of the perturbation determinant for the matrix $L(t)$ associated with the spectrum $\mu(t)$. More precisely, in approximation sense, we can first solve the discrete equilibrium measure problem \eqref{nn}, which leads to a sequence of matrices $\{L_n\}$, converging weakly \eqref{new} to the operator $L(t)$. Correspondingly, the ratios of their perturbation determinants will have a limit related to $\Psi(z, t; V)$. 

For values of $t$ corresponding to a classical LG solution, the gradient of $\Psi(z, t; V)$ will vanish precisely along the boundary $\p \Omega(t)$, since it is given by the equation $V'(z) + C_{\mu_t}(z) = 0$, where $C_{\mu_t}$ is the Cauchy transform of $\mu_t$, which coincides with $C_{\Omega(t)}$ in a neighborhood of $z = \infty$. Therefore, it is possible to define a ``weak LG boundary" by the set $\{ \p_z \Psi(z, t; V) = 0 \}$ for values of $t$ beyond $t_c$, which may consist of a union of domain boundaries and analytic arcs (regarded as the singular boundary sets for domains in the limit of empty interior). 

Along the support of $\mu_t$, $\Psi(z, t; V)$ may be assigned the value 0, as the tangential derivative vanishes. The two limit normal derivatives are equal and opposite signs. 

The relation to DLA is obtained by first formulating it as a limit process that amounts to a ``randomized" logarithmic capacity problem (and, therefore, reduced modulus, c.f. \S~\ref{sec}). Consider a measure $\mu_t$ supported on a continuum $K_t$ consisting of a finite set of analytic arcs, in a neighborhood of the origin, and the circle $K_R : = \{|z| = R\}$, such that $K_t \subset B_0(R)$ (the degenerate case corresponds to $K_t = \{ 0 \}$). Let $D_t$ denote the balayage of  $\mu_t$, and $\varphi_t: D_t^c \to \overline{\mathbb{D}}^c$ the conformal map sending $\infty$ into $\infty$, with real conformal radius. Denote by $\Sigma_t(R) : = \varphi_t(K_R)$, and $P_t$ the perimeter of  $\Sigma_t(R)$. The probability measure $\nu_t$ defined on $\p \mathbb{D}$ by $\nu_t(\gamma) := \ell(\widehat{\gamma})/P_t$, where $\ell(\widehat{\gamma})$ is the arclength measure of the arc $\widehat{\gamma}$, the intersection between $\Sigma_t(R)$ and the radial family intersecting $\gamma$, describes a ``deterministic" limit of DLA, in the sense that it obeys the same growth law, averaged over an uniform prior distribution over $K_R$. More precisely, the growth process is defined by the inverse balayage of the measure $\chi_{D_t} + \tilde \nu_t dt$, where $\tilde{\nu}_t$ is the pre-image (under $\varphi_t$) of the measure $\nu_t$, that is $\mu_{t + dt} = \mu_t + \lambda_t$, with $\lambda_t$ the inverse balayage of   $ \tilde \nu_t dt$.  

The model can be generalized by replacing the uniform measure over $\Sigma_t(R)$ by any other probability measure with support on $\Sigma_t(R)$. DLA proper is obtained in the limit $R \to \infty$, sampling with uniform probability over the circle, and replacing the deterministic flow given by the radial family of curves (equivalently, by the vector field given by the Cauchy transform of $\mu_t$, before applying the conformal map $\varphi_t$), by the It\^{o} process given by adding a Wiener process to the field $C_{\mu_t}(z)$. 

\section{Concluding remarks}

The singular perturbation nature of classical LG is the main source of complications when seeking to embed this boundary dynamics process into more general integrable structures. The weak formulation briefly described here has the advantage of connecting at once with a larger class of optimization problems for which post-critical solutions are known to exist (equilibrium measures in the sense of Chebotarev-Gonchar \cite{Gonchar:84}), as well as to free stochastic processes \cite{V} with the infinite divisibility property, which is a promising path towards formulating a quantum (maximum entropy) generalization of the classical problem. 

\section{Acknowledgments}

The author is grateful to the referees for comments and suggestions which have improved the present work.


 \bibliographystyle{amsplain}

\end{document}